\documentclass[12pt]{article}
\usepackage{graphicx}
\usepackage{subfigure}
\begin{document}
\title{
         {\Large
                 {\bf
Electromagnetic Form Factor of the $\eta$ meson in the Light Cone QCD
                 }
         }
      }

\author{\vspace*{1cm}
{\small T. M. Aliev$^a$ \thanks
{e-mail: taliev@metu.edu.tr}\,\,,
A. \"{O}zpineci$^b$ \thanks
{e-mail: ozpineci@ictp.trieste.it}\,\,,}
\\
{\small a Physics Department, Middle East Technical University, 
06531 Ankara, Turkey}\\
{\small b  The Abdus Salam International Center for Theoretical Physics,
I-34100, Trieste, Italy} }
\date{\today}

\begin{titlepage}

\maketitle
\begin{abstract}
The electromagnetic form factor of the $\eta$ meson is studied using the Light Cone QCD sum rules
approach taking in to account higher twists contributions and $SU(3)$ flavor symmetry breaking
effects.
\end{abstract}
\thispagestyle{empty}
\end{titlepage}
\section{Introduction}

The QCD sum rules \cite{R1} is one of the most powerful tools in studying the low energy hadron
physics on the basis of QCD.  
In this method, physical observables of hadrons are related with QCD
vacuum via a few condensates. 
In the literature, a new, widely discussed alternative to traditional sum rules, namely light cone
QCD sum rules (LCQSR) is a convenient tool for the study of exclusive processes which involve the emission
of a light particle.

The LCQSR is based on the operator product expansion on the light cone, which is an expansion over
the twist of the operators rather then the dimensions as the traditional QCD sum rules.
In this method all the non perturbative dynamics encoded in the so called wave functions, which
represents as the matrix element of the nonlocal operators between vacuum and the corresponding
hadron (more about this method can be found in \cite{R2,R3}). 
One of the promising ways for obtaining information about wave functions is experimental study of form factors. 
The pion electromagnetic form factor was calculated in the framework of LCQSR in \cite{R4,R5}.
Recently the electromagnetic form factors of pion and Kaon is analyzed in \cite{R6} taking into
account the effects of higher twist contributions and the $SU(3)$ flavor symmetry breaking. 

The aim of this work is the calculation of the electromagnetic form factor of the $\eta$ meson taking
into account twist $3$ and twist $4$ wave functions of the $\eta$ meson as well as the $SU(3)$
symmetry breaking effects, i.e. the mass of the strange quark and the different value of the strange
quark condensate.

The paper is organized as follows: In Sect. 2, we derive the sum rules for the electromagnetic
form factor of the $\eta$ meson which is then analyzed numerically in section 3. Summary of our
results and conclusions are presented at the end of Sect. 3.

\section{Sum Rules for the $\eta$ Meson Electromagnetic Form Factor}

In this section we calculate the e.m. form factor of the $\eta$ meson employing LCQSR.
For this purpose, we consider the following correlation function:
\begin{eqnarray}
\Pi_{\mu \nu} (p,q) = 
i \int d^4 x \langle \eta(p) \vert  j_\mu^{el}(x) j_\nu^\eta(0) \vert 0 \rangle 
\label{eq1}
\end{eqnarray}
where
\begin{eqnarray}
j_\nu^\eta = \frac{1}{\sqrt6} \left(\bar u \gamma_\nu \gamma_5 u + \bar d \gamma_\nu \gamma_5 d -2 \bar s
\gamma_\nu \gamma_5 s \right)
\end{eqnarray}
is the $\eta$ meson interpolating current and
\begin{eqnarray}
j_\mu^{el} = \sum_{q=u,d,s} e_q \bar q \gamma_\mu q 
\end{eqnarray}
is the electromagnetic current.

Here, we would like to remark that in this work we will follow the usual parametrization of the
$\eta-\eta'$ mixing in octet-singlet basis and due to the smallness of the mixing angles (see e.g.
\cite{R7}), they will be neglected. 

Saturating Eq. (\ref{eq1}) with the $\eta$ meson state, for the phenomenological part of the
correlator, we get:
\begin{eqnarray}
\Pi_{\mu \nu} = \frac{
\langle \eta(p) \vert j_\mu^{el} \vert \eta(p+q) \rangle \langle \eta(p+q) \vert j_\nu^\eta
\vert 0 \rangle}{(p+q)^2 - m_\eta^2}
\label{eq2}
\end{eqnarray}
The matrix elements  entering Eq. (\ref{eq2}) are defined as:
\begin{eqnarray}
\langle \eta(p) \vert j_\mu^{el} \vert \eta(q+p) \rangle &=& F_\eta^{el} (2 p +q )_\mu 
\nonumber \\
\langle \eta(p+q) \vert j_\nu^\eta \vert 0 \rangle &=& -i f_\eta (p+q)_\nu
\label{eq3}
\end{eqnarray}
Putting Eq. (\ref{eq3}) to (\ref{eq2}), for the phenomenological part of the correlator, Eq.
(\ref{eq1}), we get:
\begin{eqnarray}
\Pi_{\mu \nu}(p,q) = - i f_\eta F_\eta^{el} \frac{
2 p_\mu p_\nu + 2 p_\mu q_\nu + p_\nu q_\mu + q_\mu q_\nu}{
(p+q)^2 - m_\eta^2}
\label{eq4}
\end{eqnarray}

From Eq. (\ref{eq4}), it is seen that there are four different structures that can be used to
extract the form factor. In this work we will concentrate only on the $p_\mu p_\nu$ and $q_\mu q_\nu$
structures. Hence one can write
\begin{eqnarray}
\Pi_{\mu \nu}(p,q) = \Pi_1(q^2,(p+q)^2) p_\mu q_\nu + \Pi_2(q^2,(p+q)^2) q_\mu q_\nu + \ldots
\end{eqnarray}
where $\ldots$ stand for the remaining structures and
\begin{eqnarray}
\Pi_1(q^2,(p+q)^2) &=& -2 i f_\eta \frac{F_\eta^{el}(Q^2)}{(p+q)^2 - m_\eta^2} 
\nonumber \\
\Pi_2(q^2,(p+q)^2) &=& - i f_\eta \frac{F_\eta^{el}(Q^2)}{(p+q)^2 - m_\eta^2}
\label{eq8}
\end{eqnarray}
where $Q^2= - q^2$.
For the invariant amplitudes $\Pi_i(q^2,(p+q)^2)$, one can write a general dispersion relation in
$(p+q)^2$ in the form:
\begin{eqnarray}
\Pi_i(q^2,(p+q)^2) = \int ds \frac{\rho_i(s,q^2)}{s - (p+q)^2} + ~subtraction~terms
\end{eqnarray}
where the spectral functions corresponding to Eq. (\ref{eq8}) are 
\begin{eqnarray}
\rho_1(s,q^2) = 2 \rho_2(s,q^2) = 2 i F_\eta^{el}(Q^2) \delta(s - m_\eta^2)
\end{eqnarray}

Our next task is the calculation of the correlator from the QCD side. 
If the $(p+q)^2$ and $q^2$ are large and space like, the OPE in Eq. 
(\ref{eq1}) can be expanded near the light cone in terms of the $\eta$ 
meson wave functions, i.e. the matrix elements of the nonlocal operators between vacuum and the
$\eta$ meson, like $\langle 0 \vert  \bar q \Gamma_i q \vert 0 \rangle$ where $\Gamma_i$ is an
arbitrary Dirac matrix. 
As we already noted that we will take into account twist 3 and twist 4 wave functions beyond 
the leading twist and keep terms up to linear order in the strange quark mass.
Up to twist 4, the $\eta$ meson wave functions are defined in the following way \cite{R8}:
\begin{eqnarray}
&&\langle \eta (p) \vert \bar q(x) \gamma_\mu \gamma_5 q(0) \vert 0 \rangle =
-i f_\eta p_\mu \int_0^1 du e^{i u px} \left[ \varphi_\eta(u) + \frac{m_\eta^2 x^2}{16} A(u) \right]
\nonumber \\ &&
- \frac{i}{2} f_\eta m_\eta^2 \frac{x_\mu}{px} \int_0^1 du e^{i u px} B(u)
\nonumber \\ &&
\langle \eta (p) \vert \bar q(x) i \gamma_5 q(0) \vert 0 \rangle =
f_\eta \mu_\eta \int_0^1 du e^{i u px} \varphi_P(u)
\nonumber \\&&
\langle \eta (p) \vert \bar q(x) \sigma_{\mu \nu} \gamma_5 q(0) \vert 0 \rangle =
\nonumber \\ &&
i \frac{f_\eta \mu_\eta}{6} \left( 1 - \frac{\mu_\eta^2}{m_\eta^2 f_\eta^2} \right)
\left[ p_\alpha x_\beta - p_\beta x_\alpha \right] \int_0^1 du e^{i u p x} \varphi_\sigma(u)
\nonumber \\&&
\langle \eta(p) \vert \bar q(x) \sigma_{\alpha \beta} \gamma_5 G_{\mu \nu}(ux) q(0) \vert 0
\rangle =
\nonumber \\ &&
i f_{3 \eta} \left[ p_\alpha \left( p_\mu g_{\nu \beta} - p_\nu g_{\mu \beta} \right)
- p_\beta \left(p_\mu g_{\nu \alpha} - p_\nu g_{\mu \alpha} \right) \right] 
\int {\cal D} \alpha_i \varphi_{3 \eta} (\alpha_i) e^{i p x (\alpha_1 + u \alpha_3)}
\nonumber \\&&
\langle \eta(p) \vert \bar q(x) \gamma_\mu \gamma_5 g_s G_{\alpha \beta} (ux) q(0) \vert 0
\rangle
\nonumber \\ &&
f_\eta m_\eta^2 \left[ p_\beta \left( g_{\alpha \mu} - \frac{x_\alpha p_\beta}{px} \right)
- p_\alpha \left( g_{\beta \mu} - \frac{x_\beta p_\mu}{px} \right) \right] 
\int {\cal D} \alpha_i \varphi_\perp (\alpha_i) e^{i p x (\alpha_1 + u \alpha_3)}
\nonumber \\ &&
+ f_\eta m_\eta^2 \frac{p_\mu}{px} \left( p_\alpha x_\beta - p_\beta x_\alpha \right)
\int {\cal D} \alpha_i \varphi_\parallel (\alpha_i) e^{i px (\alpha_1 + u \alpha_3)}
\nonumber \\&&
\langle \eta(p) \vert \bar q (x) \gamma_\mu \tilde G_{\alpha \beta} (u x) q(0) \vert 0
\rangle = 
\nonumber \\ &&
i f_\eta m_\eta^2 \left[ p_\beta \left( g_{\alpha \mu} - \frac{x_\alpha p_\beta}{px} \right)
- p_\alpha \left( g_{\beta \mu} - \frac{x_\beta p_\mu}{px} \right) \right] 
\int {\cal D}\alpha_i \tilde \varphi_\perp (\alpha_i) e^{i px (\alpha_1 + u \alpha_3)} 
\nonumber \\ &&
i f_\eta m_\eta^2 \frac{p_\mu}{px} \left( p_\alpha x_\beta - p_\beta x_\alpha \right)
\int {\cal D} \alpha_i \tilde \varphi_\parallel(\alpha_i) e^{i px(\alpha_i + u \alpha_3)}
\end{eqnarray}

After contracting quark fields, the correlator function becomes:
\begin{eqnarray}
\Pi_{\mu \nu} = \sum_{i=u, d, s} e_i c_i \int d^4 x e^{i p x} \langle
\eta (p) \vert \bar q_i(x) S_i(x) q_i(0) + \bar q_i(0) S_i (-x) q_i(x) \vert 0 \rangle
\end{eqnarray}
where $S_i(x)$ is the full propagator of the light quark, and $c_{u,d} = 1$ and $c_s = -2$.
During the analysis, the following full light quark propagator is used:
\begin{eqnarray}
S(x) &=& \frac{i \not\!x}{2\pi^2x^4} - \frac{m_q}{4 \pi^2 x^2} - \frac{\langle \bar q q \rangle}{12}
\left(1 - i \frac{m_q}{4} \not\!x \right) - \frac{x^2}{192} m_0^2 \langle \bar q q \rangle 
\left( 1 - i \frac{m_q}{6}\not\!x \right) 
\nonumber \\ &&
- i g_s \int_0^1 du \left[\frac{\not\!x}{16 \pi^2 x^2} G_{\mu \nu} (ux) \sigma_{\mu \nu} - u x^\mu
G_{\mu \nu} (ux) \gamma^\nu \frac{i}{4 \pi^2 x^2} 
\right. \nonumber \\ && \left.
- i \frac{m_q}{32 \pi^2} G_{\mu \nu} \sigma^{\mu
\nu} \left( \ln \left( \frac{-x^2 \Lambda^2}{4} \right) + 2 \gamma_E \right) \right]
\end{eqnarray}
where $\Lambda$ is the energy cut off separating perturbative and non perturbative regimes. 

In order to suppress the contributions of higher states and continuum and also to eliminate the
subtraction terms in the dispersion relation, the result is Borel transformed with respect to
$-(p+q)^2$
After lengthy calculation for the theoretical part of the correlator function, we get the following
result
\begin{eqnarray}
&&\sqrt6 \Pi_{\mu \nu} (Q^2,M^2) =
\nonumber \\ &&
 2 (e_d + e_u - 2 e_s) \frac{f_\eta^2 m_\eta^2}{M^2}\int_0^1 du \int_0^u dk 
\frac{e^{-\frac{s(k)}{M^2}}}{k^2}  \left[ B(u) + B(\bar u) \right]
\left( k p_\mu + q_\mu \right) \left( k p_\nu + q_\nu \right) 
\nonumber \\ &&  
+ \int_0^1 du \frac{e^{-\frac{s(u)}{M^2}}}{u^2} \left\{
-\frac{2 m_s e_s}{3} 
\left( 1 - \frac{\mu_\eta^2}{m_\eta^2 f_\eta^2} \right) \frac{\mu_\eta}{M^2} \left[ \varphi_\sigma(u) - \varphi_\sigma(\bar u)  \right]
\left( p_\nu q_\mu - p_\mu q_\nu \right) \right.
\nonumber \\ &&  
+ 2 m_s e_s f_\eta \mu_\eta u \left[ \varphi_P(u) - \varphi_P(\bar u) \right] g_{\mu \nu}
\nonumber \\ && 
-\frac{1}{2} (e_d + e_u - 2 e_s) f_\eta  
\left[ \varphi_\eta(u) - \varphi_\eta(\bar u) \right]
\nonumber \\ && 
\times \left( 4 u^2 p_\mu p_\nu + 2 u (p_\nu q_\mu +  p_\mu q_\nu) - (Q^2 + m_\eta^2 u^2) g_{\mu \nu}
\right) 
\nonumber \\ &&
+ \frac{1}{8} (e_d + e_u - 2 e_s) \frac{f_\eta m_\eta^2}{u M^2} 
 \left[ A(u) - A(\bar u) \right] 
\nonumber \\ && \left. 
\times \left( 4 u^2 p_\mu p_\nu + 2 u (p_\nu q_\mu + p_\mu q_\nu) + (Q^2 - M^2 u + m_\eta^2 u^2) g_{\mu
\nu} \right) \right\}
\nonumber \\ &&
+ \int_0^1 dv \int {\cal D}\alpha_i 
\frac{e^{- \frac{s(\alpha_1 + v \alpha_3)}{M^2}}}{(\alpha_1 + v \alpha_3)^2}
\left \{ - 4 (e_d + e_u - 2 e_s) \frac{f_\eta m_\eta^2}{M^2} \left[ {\cal A}_\perp (\alpha_i) + {\cal
A}_\perp(\bar \alpha_i) \right]  
\right.
\nonumber \\ && 
\times \left[ (v - \bar v) (\alpha_1 + v \alpha_3) p_\mu p_\mu + v p_\mu q_\nu - \bar v p_\nu q_\mu \right]
\nonumber \\ &&
-\frac{1}{2} (e_d + e_u - 2 e_s) \frac{f_\eta m_\eta^2}{M^2} 
\frac{1}{(\alpha_1 + v \alpha_3)} 
\nonumber \\ &&
\times \left[ (v - \bar v) \left( {\cal A}_\parallel (\alpha_i) + {\cal A}_\parallel(\bar
\alpha_i) \right) + \left( {\cal V}_\parallel (\alpha_i) + {\cal V}_\parallel(\bar\alpha_i) \right) \right]
\nonumber \\ &&
\times \left[ 4 (\alpha_1 + v \alpha_3)^2 p_\mu p_\mu +2 (\alpha_1 + v \alpha_3) (p_\nu q_\mu + p_\mu q_\nu) 
\right. \nonumber \\ && \left.  
- \left(Q^2 + (\alpha_1 + v \alpha_3)^2 m_\eta^2 - (\alpha_1 + v \alpha_3) M^2\right) g_{\mu
\nu} \right] 
\nonumber \\ &&
- 4 e_s m_s \frac{f_\eta \mu_\eta^2}{M^2} 
\left[ v {\cal T}(\alpha_i) - \bar v {\cal T}(\bar \alpha_i) \right]
\left( 4 p_\mu p_\nu - m_\eta^2 g_{\mu \nu} \right)
\nonumber \\ && 
\left. + 4 (e_d + e_u - 2 e_s)  \frac{f_\eta m_\eta^2}{M^2}
\left[ {\cal V}_\perp(\alpha_i) - {\cal V}_\perp(\bar \alpha_i) \right]
\left[ (\alpha_1 + v \alpha_3) p_\mu p_\nu + v p_\mu q_\nu + \bar v p_\nu q_\mu \right] \right\} 
\nonumber \\
\label{eq14}
\end{eqnarray}
where $M^2$ is the Borel parameter and
\begin{eqnarray}
s(k) = \frac{\bar k}{k} \left( m_\eta^2 k + Q^2 \right)
\nonumber 
\end{eqnarray}
where $Q^2= - q^2$.
The contributions of the higher states and continuum are
subtracted using the quark hadron duality (The details can be found in \cite{R9}).
Hence for the invariant functions in Eq. (\ref{eq8}), we get the following results:
\begin{eqnarray}
\sqrt{6} \Pi_1(Q^2,M^2) &=& \int_0^1 du e^{-\frac{s(u)}{M^2}} \left\{ 
\frac{1}{2} (e_d + e_u - 2 e_s) \frac{f_\eta m_\eta^2}{u M^2} 
\left( A(u) - A(\bar u) \right) \right.
\nonumber \\ &&
+ 2 (e_d + e_u - 2 e_s) \frac{f_\eta m_\eta^2}{M^2} 
u \left( B(u) + B(\bar u) \right) 
\nonumber \\ &&
- 2 (e_d + e_u - 2 e_s) f_\eta 
\left. \left( \varphi_\eta(u) - \varphi_\eta(\bar u) \right) \right\}
\nonumber \\ &&
+ \int_0^1 dv \int {\cal D}\alpha_i  e^{- \frac{s(\alpha_1 + v \alpha_3)}{M^2}}\left\{
- 2 (e_u + e_d - 2 e_s) \frac{f_\eta m_\eta^2}{M^2}  
\frac{1}{\alpha_1 + v \alpha_3}  \right.
\nonumber \\ &&
\times
\left[(v- \bar v)
\left( {\cal A}_\parallel(\alpha_i) + {\cal A}_\parallel(\bar \alpha_i) +
2 {\cal A}_\perp(\alpha_i) + 2 {\cal A}_\perp(\bar \alpha_i) \right) 
\right. \nonumber \\ && \left.
- {\cal V}_\parallel(\alpha_i) + {\cal V}_\parallel(\bar \alpha_i) -
2 {\cal V}_\perp(\alpha_i) + 2 {\cal V}_\perp(\bar \alpha_i) \right]
\nonumber \\ &&
+ 16 m_s e_s \frac{f_\eta \mu_\eta}{M^2} \left.\frac{1}{(\alpha_1 + v \alpha_3)^2} 
 \left(v {\cal T}(\alpha_i) - \bar v {\cal T}(\bar \alpha_i) \right) \right\}
 \label{eq15}
 \\
\sqrt6 \Pi_2(q^2,M^2) &=& 2 (e_d + e_u - 2 e_s) \frac{f_\eta m_\eta^2}{M^2} \int_0^1 du \int_0^u dk
\frac{e^{-\frac{s(k)}{M2}}}{k^2}  \left[ B(u) + B(\bar u) \right]
\label{eq16}
\end{eqnarray}
\section{Numerical Analysis}
In this section, we will carry out the numerical analysis our results.
We used the following values of the constants appearing in Eqs. (\ref{eq15}) and (\ref{eq16})
\cite{R8}:
$m_s(1~GeV)=150~MeV$, $f_\eta(1~GeV) = 0.156~GeV$, and $\mu_\eta(1~GeV) = 0.034~GeV^2$. 

%
%
In numerical analysis, we systematically take into account all ${\cal O}(m_s) \sim {\cal
O}(m_\eta^2)$ effects, which have following sources: a) from the correlator when we take $p^2 =
m_\eta^2$; b) from $s$-quark propagator, where terms linear in $m_s$ are retained; c) $SU(3)$ flavor
symmetry breaking effects which can appear in three different ways: i)  via $SU(3)$ violation in
normalization factors, i.e. $f_\eta \neq f_\pi$; ii) Meson mass corrections to the twist four wave
functions; iii) via an asymmetry of the wave function with respect to the interchange of quark and
antiquark.
At leading twist two wave function, this leads to the appearance of non vanishing odd coefficients
of the Gegenbauer expansion, i.e.
\begin{eqnarray}
\varphi_\eta(u,\mu) = 6 u \bar u \left[ 1 + \sum_{n=1} a_n(\mu) C_n^{3/2}(2 u - 1) \right]
\nonumber 
\end{eqnarray}
where $C_n^{3/2}(2 u -1)$ are Gegenbauer polynomials. In our calculation, we only considered
$a_1(\mu)$ and $a_2(\mu)$

From Eq. (\ref{eq15}), it follows that leading twist two wave function appear only in the form
$\varphi(u) - \varphi(\bar u)$, i.e. they contribute as a result of pure $SU(3)$ violating effects.
In the numerical calculation, we kept the terms $\propto a_1(\mu)$ only. The first coefficient
$a_1(\mu)$ was calculated in \cite{R10} for $K$ and $K^*$ mesons. In the present work we have
estimated $a_1(\mu)$ for the $\eta$ meson and found that $a_1(\mu) = - 0.20 \pm 0.02$. The ${\cal
O}(\alpha_s)$ corrections to the leading twist two wave function has been calculated in \cite{R5} in
the chiral limit and this correction is also included in our calculations.

\begin{figure}[h!]
$\left. \right.$
\begin{center}
\subfigure[]{
\includegraphics[width=5.6cm,angle=-90]{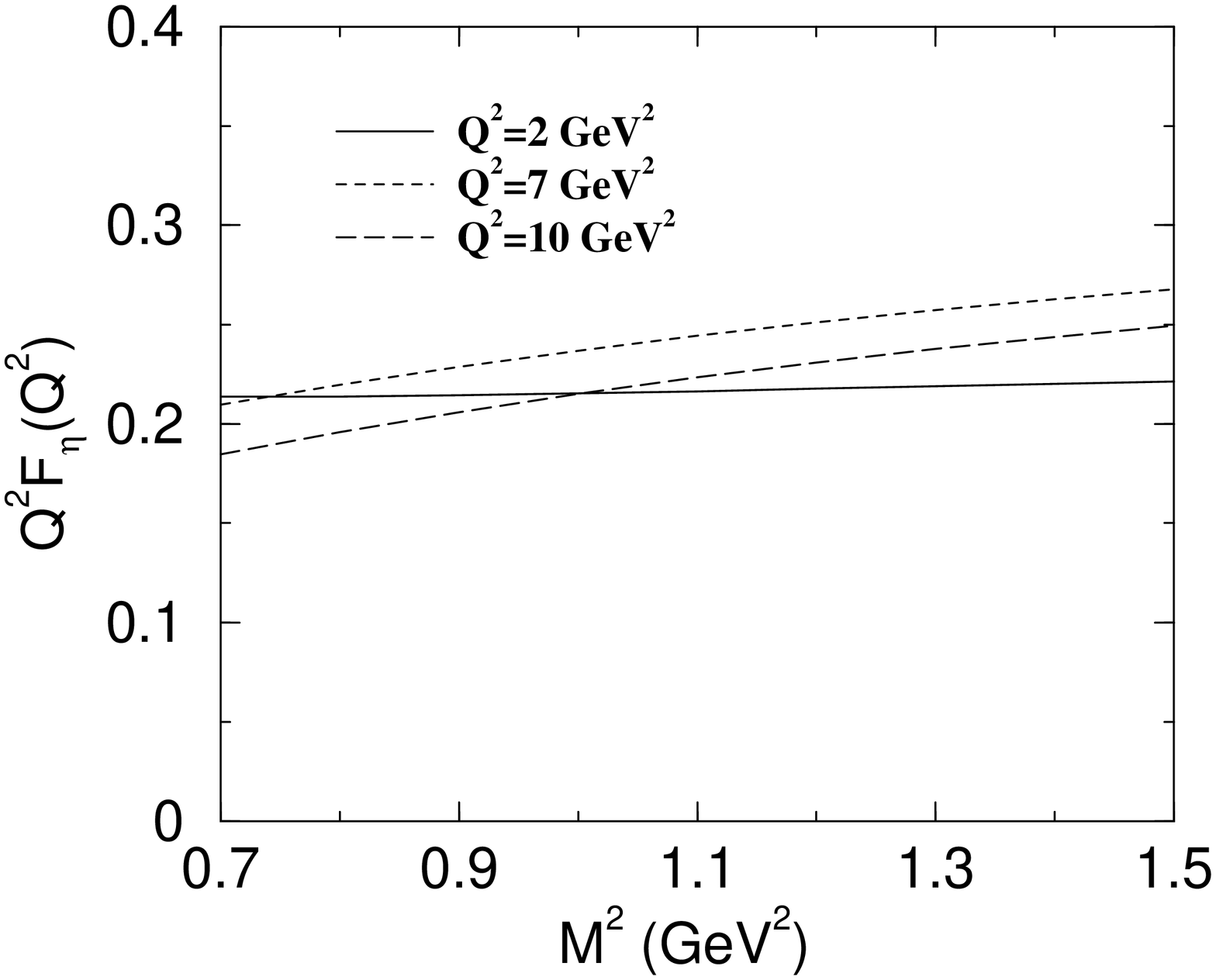}}
\subfigure[]{
\includegraphics[width=5.6cm,angle=-90]{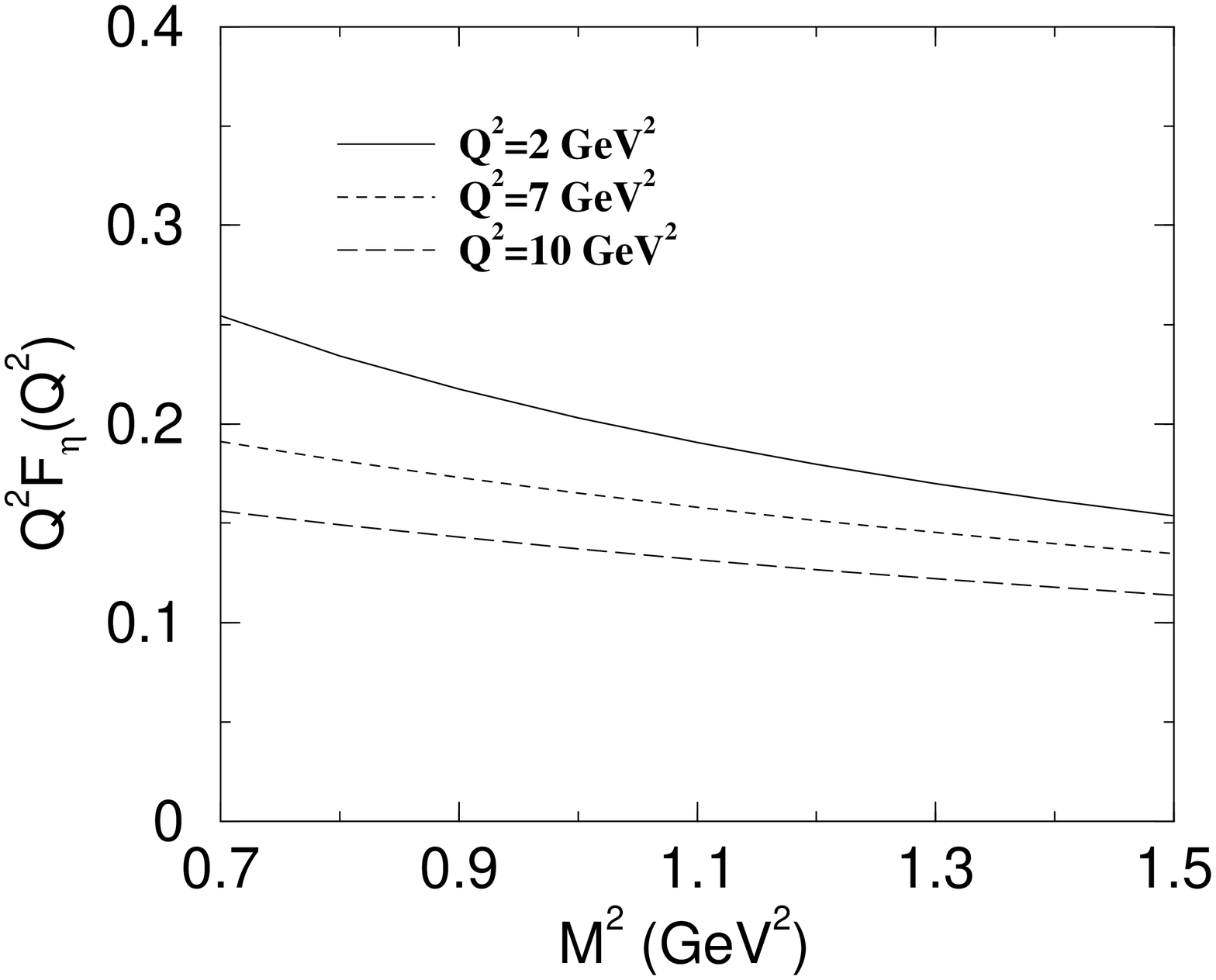}}
\end{center}
\vspace{-0.5cm}
\caption{The dependence of $Q^2 F_\eta(Q^2)$ on the Borel parameter at $Q^2=2,~7,~10~GeV^2$ for the
(a) $p_\mu p_\nu$ and (b) $q_\mu q_\nu$ structures}
\label{msq_s0_1.5}
\end{figure}

\begin{figure}[h!]
$\left. \right.$
\begin{center}
\subfigure[]{
\includegraphics[width=5.6cm,angle=-90]{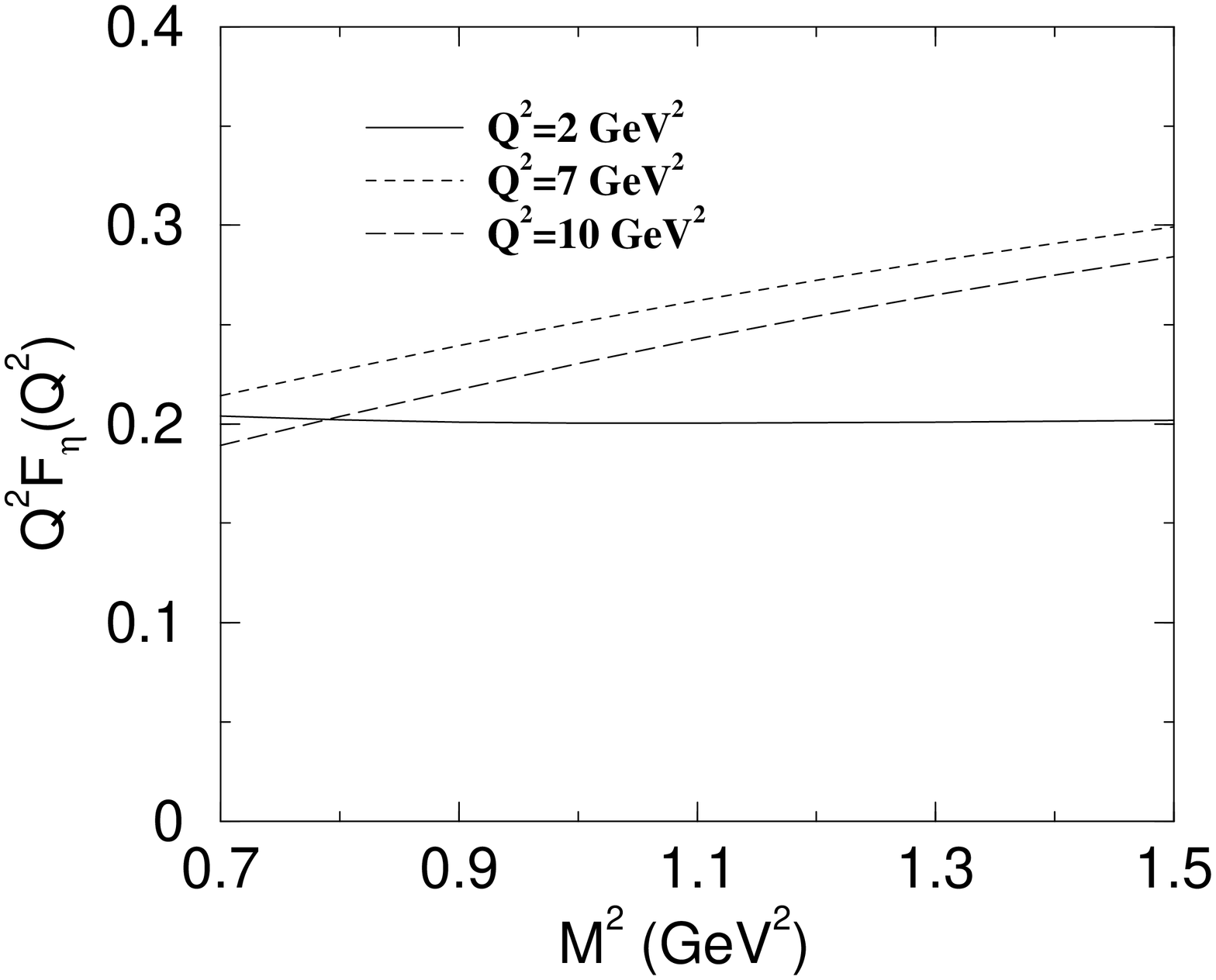}}
\subfigure[]{
\includegraphics[width=5.6cm,angle=-90]{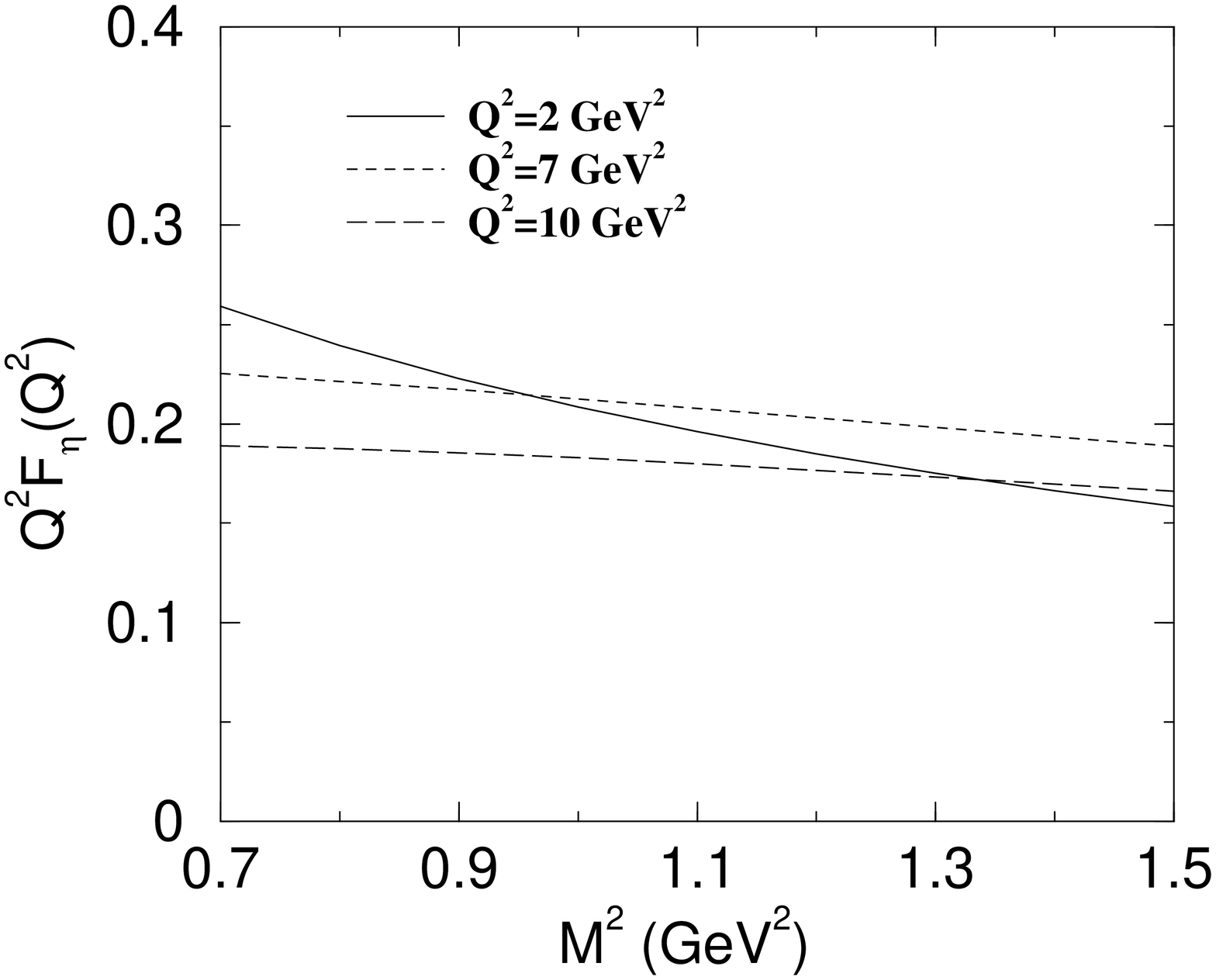}}
\end{center}
\vspace{-0.5cm}
\caption{The same as Fig. (\ref{msq_s0_1.5}) but at $s_0=2.5$}
\label{msq_s0_2.5}
\end{figure}

For the analysis of the sum rules, one should determine a value of the continuum threshold and a
region of the Borel parameter $M^2$ in which the predictions do not depend on the value of $M^2$.
From the analysis of the mass sum rules for $\eta$, the continuum threshold $s_0$ is found to be
around $s_0 \simeq 2.0~GeV^2$. In our analysis we will take $s_0=1.5~GeV^2$ and $s_0=2.5~GeV^2$ to
study the dependence of our predictions on $s_0$. In choosing the suitable region for the Borel
parameter $M^2$, one should keep in mind the $M^2$ should be big enough to suppress the
contributions of higher dimensional operators in the operator product expansion of the correlator
function, and at the same time small enough to have a considerable suppression of the contributions
of the higher states and the continuum. In our analysis, we will take $0.7 < M^2 < 1.5$ in which
region the contributions of both the higher dimensional operators and the contributions of the
higher states and continuum are less then $30\%$. In Figs. (\ref{msq_s0_1.5}) and
(\ref{msq_s0_2.5}), we plot the dependence of $Q^2 F_\eta(Q^2)$ on the Borel parameter $M^2$ for
$Q^2=2,~7,~10~GeV^2$ and for the values of the continuum threshold $s_0=1.5,~2.5~GeV^2$.
From the figures one sees that the predictions from the $p_\mu p_\nu$ structure are less stable for
larger values of $Q^2$, which is reasonable as the sum rules is expected to break for very large
values of $Q^2$. For smaller values of $Q^2$, the predictions are almost independent of the exact
value of both $s_0$ and $M^2$.

\begin{figure}[h!]
$\left. \right.$
\begin{center}
\includegraphics[width=7cm,angle=-90]{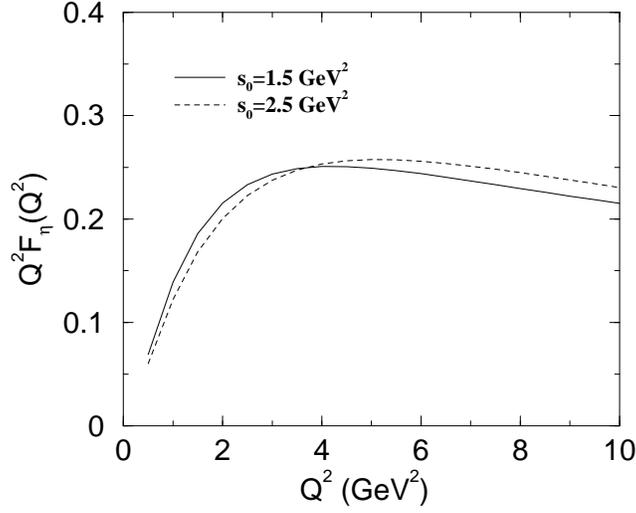}
\end{center}
\vspace{-0.5cm}
\caption{The dependence of $Q^2 F_\eta(Q^2)$ on the $Q^2$ obtained from the structure $p_\mu p_\nu$
at $M^2=1~GeV^2$ and at $s_0=1.5,~2.5~GeV^2$} 
\label{qsq1_msq_1}
\end{figure}

\begin{figure}[h!]
$\left. \right.$
\begin{center}
\includegraphics[width=7cm,angle=-90]{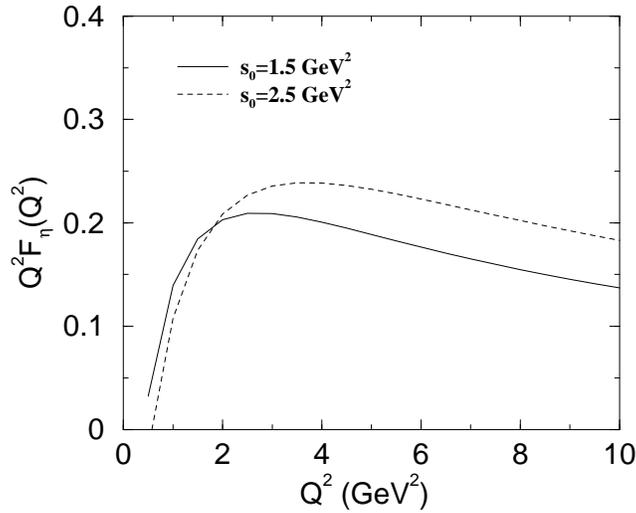}
\end{center}
\vspace{-0.5cm}
\caption{The same as Fig. (\ref{qsq1_msq_1}) but for the structure $q_\mu q_\nu$ }
\label{qsq4_msq_1}
\end{figure}

In Figs. (\ref{qsq1_msq_1}) and (\ref{qsq4_msq_1}), the dependence of $Q^2 F_\eta(Q^2)$ on $Q^2$ is
depicted at the value of the Borel parameter $M^2=1~GeV$ and for two different values of the
continuum threshold $s_0=1.5$ and $2.5~GeV$. As can be seen also in these figures, the predictions
obtained from the structure $p_\mu p_\nu$ is more robust to variations in the continuum threshold
$s_0$ then the predictions obtained from the $q_\mu q_\nu$ structure. A fit to the form factor
$F_\eta(Q^2)$ can be written in the form:
\begin{eqnarray}
F_\eta(Q^2) = \frac{F_\eta(0)}{1 + b_1 \frac{Q^2}{m_\eta^2} + b_2 \left( \frac{Q^2}{m_\eta^2} \right)^2}
\end{eqnarray}
From the analysis of the sum rules obtained from the $p_\mu p_\nu$ structures, for the constants
$a_i$ we obtained the following values:
$F_\eta(0) = 0.17 \pm 0.02$, $b_1 = (8 \pm 2) 10^{-2}$, and $b_2 =(4 \pm 1) 10^{-3}$.

\section*{Acknowledgement}
One of the autors (T.M.A) sincerely thanks the A. Salam International Centre for Theoretical 
Physics and High Energy section, where essential part of this work is done,  and its head Prof.S.Randjbar-Daemi,for 
hospitality and financial support.

\newpage

\end{document}